\documentclass[10pt, journal,compsoc]{IEEEtran}
\usepackage{cite}
\usepackage[T1]{fontenc}
\usepackage{amsmath}
\usepackage{color}
\usepackage{cite} 
\usepackage{array}
\usepackage[caption=false,font=footnotesize]{subfig}
\usepackage{stfloats}
\usepackage{url}
\usepackage{multirow}
\usepackage{amssymb}
\usepackage{textcomp}
\usepackage{amsthm}
\usepackage{algorithm}          
\usepackage{algorithmicx}     
\usepackage{algpseudocode}  
\usepackage{setspace}
\usepackage{float}
\usepackage{makecell}
\usepackage{amsfonts}
\usepackage{algpseudocode}
\usepackage{bm}
\usepackage{booktabs}
\usepackage{mathbbol}
\usepackage{lipsum}
\usepackage{upgreek}
\usepackage{graphicx}
\usepackage{tabularx}

\makeatletter
\newenvironment{breakablealgorithm}
{
	\begin{center}
		\refstepcounter{algorithm}
		\hrule height.8pt depth0pt \kern2pt
		\renewcommand{\caption}[2][\relax]{
			{\raggedright\textbf{\ALG@name~\thealgorithm} ##2\par}%
			\ifx\relax##1\relax 
			\addcontentsline{loa}{algorithm}{\protect\numberline{\thealgorithm}##2}%
			\else 
			\addcontentsline{loa}{algorithm}{\protect\numberline{\thealgorithm}##1}%
			\fi
			\kern2pt\hrule\kern2pt
		}
	}{
		\kern2pt\hrule\relax
	\end{center}
}
\makeatother

\begin{document}

\title{Distributed Traffic Engineering in Hybrid Software Defined Networks: A Multi-agent Reinforcement Learning Framework}

\author{Yingya~Guo,
        Qi~Tang,
        Yulong~Ma,
        Han~Tian
        and Kai~Chen

\IEEEcompsocitemizethanks{\IEEEcompsocthanksitem Yingya Guo, Qi Tang, Yulong Ma are with the College of Computer and Data Science, Fuzhou University; the Fujian Provincial Key Laboratory of Network Computing and Intelligent Information Processing, Fuzhou University, and also with the Key Laboratory of Spatial Data Mining \& Information Sharing, Ministry of Education Fujian, P.R.China, 350003.
\IEEEcompsocthanksitem Han~Tian and Kai~Chen are with the Department of Computer Science \& Engineering, Hong Kong University of Science \& Technology.}
}

\maketitle
\begin{abstract}
Traffic Engineering (TE) is an efficient technique to balance network flows and thus improves the performance of a hybrid Software Defined Network (SDN). Previous TE solutions mainly leverage heuristic algorithms to centrally optimize link weight setting or traffic splitting ratios under the static traffic demand. Note that as the network scale becomes larger and network management gains more complexity, it is notably that the centralized TE methods suffer from a high computation overhead and a long reaction time to optimize routing of flows when the network traffic demand dynamically fluctuates or network failures happen. To enable adaptive and efficient routing in TE, we propose a Multi-agent Reinforcement Learning method CMRL that divides the routing optimization of a large network into multiple small-scale routing decision-making problems. To coordinate the multiple agents for achieving a global optimization goal, we construct an interactive environment for training the routing agents that own partial link utilization observations. To optimize credit assignment of multi-agent, we introduce the difference reward assignment mechanism for encouraging agents to take better action. Extensive simulations conducted on the real traffic traces demonstrate the superiority of CMRL in improving TE performance, especially when traffic demands change or network failures happen.\par
\end{abstract}


%

\section{Introduction}\label{s1}

Due to the explosive growth of Internet traffic, Traffic Engineering (TE) has gained increasing attentions in achieving traffic balancing and improving network performance \cite{xiao2021leveraging}. Nowadays, the TE performance of traditional distributed network is largely constrained by its adopted shortest-path-routing protocols. Fortunately, with the emergence of Software Defined Network (SDN) \cite{xia2014survey} architecture, the decoupling of control plane and data plane enables TE to design flexible solutions for better optimizing traffic routing. However, the full-SDN enabled network, which upgrades all legacy routers with SDN switches, encounters economical and technical problems \cite{vissicchio2014opportunities}. Therefore, hybrid SDN, in which SDN switches are partially deployed in the traditional distributed networks, is widely adopted by Internet Service Providers (ISPs) as a practical solution to realize a smarter TE. Many studies have shown that TE in the hybrid SDN can achieve the network performance close to the full-SDN enabled network \cite{agarwal2013traffic}.

Previous TE solutions for hybrid SDN mainly focused on developing various heuristics \cite{agarwal2013traffic,guo2014traffic,guo2019sote,guo2019joint}. These heuristics are often human-designed and optimize routing policies only on a single traffic demand. As a result, the routing policies derived from these heuristics inevitably suffer a performance degradation for the inability to adapt to the dynamically-changing network environment, e.g. fluctuating traffic demands or network link failures. In addition, due to the high computation and communication overhead, it is impractical for these heuristics to promptly re-calculate and deploy the appropriate routing strategy in a dynamically-changing network environment.\par

As an essential branch of machine learning, Reinforcement Learning (RL) has exhibited great potential in tackling dynamic decision-making problems by enabling an experience-driven and model-free control \cite{lillicrap2015continuous}. Instead of the human-designed heuristics, RL learns an intelligent agent to adaptively and rapidly derive optimal policies according to different environments. Without any supervised information, RL can automatically accumulate a large number of valuable experience by repeatedly interacting with a virtual environment in a trial and error manner. The accumulated experience helps the intelligent agent to discover the hidden patterns in the historical data, and establish the direct relationship between dynamic environments and optimal policies.

\begin{figure}[!t]
	\begin{center}
		\includegraphics[width=\linewidth]{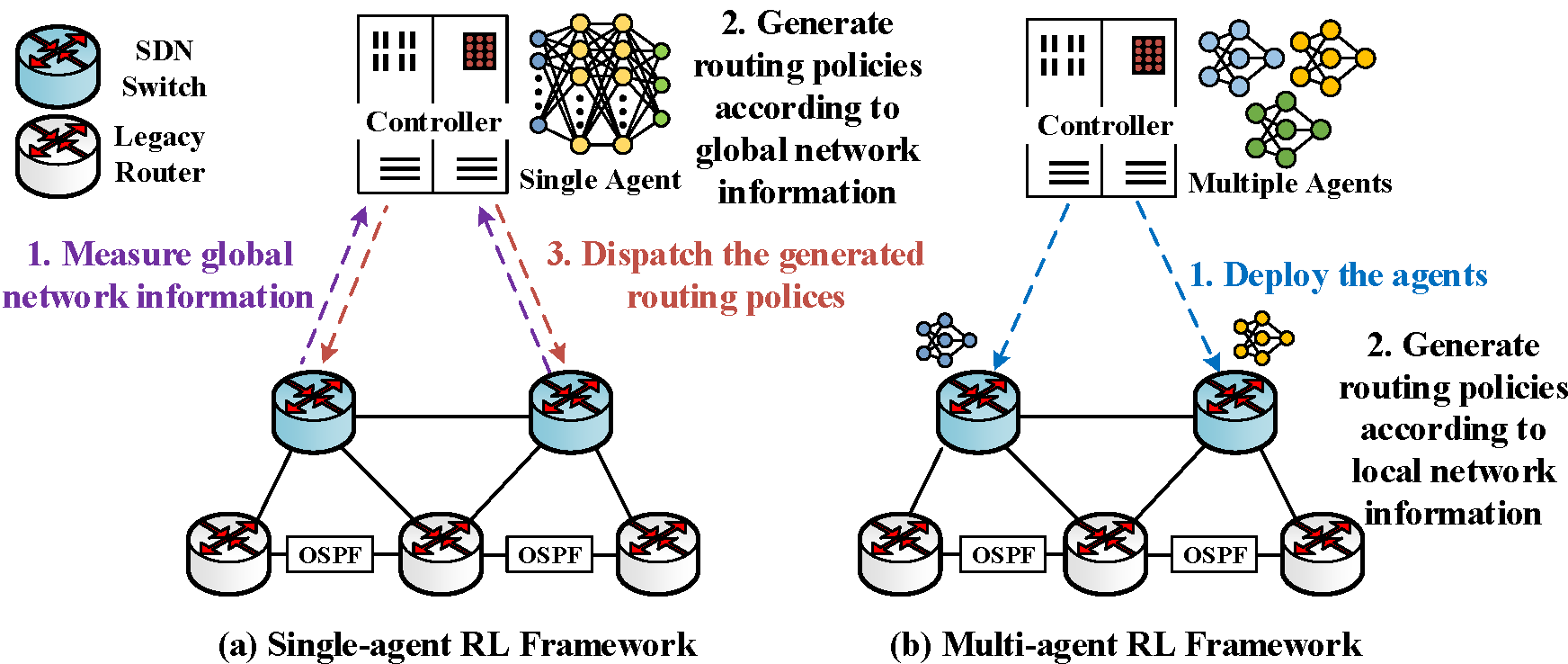}
		\caption{The illustration of different workflows of single-agent RL framework and multi-agent RL framework for achieving TE in hybrid SDN.}
		\label{workflow}
	\end{center}
\vspace{-0.3cm}
\end{figure}

Some pioneering studies have attempted to leverage RL technique to address the TE problems in the hybrid SDN \cite{tian2020traffic,guo2021traffic}. However, as the network scales increase, the action space increases rapidly and it will be intractable to make accurate online routing inference with a single agent \cite{xu2019evaluating}. Unlike these studies adopting the single-agent RL framework to implement a centralized TE, this paper proposes a multi-agent RL framework for achieving a distributed TE in a hybrid SDN. As shown in Fig. \ref{workflow}, the workflow of our proposed multi-agent RL framework exhibits two main advantages: 1) the multi-agent RL framework needs no additional communication overhead to exchange the network information when deciding the routing strategy, since it only needs the local network information rather than the global network information, which makes the routing inference more efficient; 2) the multi-agent RL framework performs a higher potential to improve network performance and exhibits better scalability, because it decreases the problem complexity and improves the convergence by decomposing a large-scale problem into several small-scale sub-problems, each of which is solved by a relatively simple and independent agent.

However, developing an efficient multi-agent RL framework to achieve distributed TE in hybrid SDN still encounters the following challenges. First, the state, reward and action functions should be carefully designed to enable an efficient multi-agent training. Constructing an interactive environment for enabling multi-agent to collaboratively learn the map between the traffic demands and routing policies poses a great challenge. Second, in a cooperative setting, the global reward generated by the joint action makes it hard to quantify the contribution of each agent and the individual reward for each agent can hardly motivate it for taking better action. Assigning reasonable credits for different agents and designing a difference reward assignment mechanism to the agents for encouraging the better actions of agents pose another challenge. \par

To address the above challenges, we innovatively propose a Counterfactual-based Multi-agent Reinforcement Learning method CMRL for improving TE performance in a hybrid SDN. Specifically, we first construct an interactive environment with carefully designed state, action and reward functions. Then, we propose a Deep Deterministic Policy Gradient (DDPG)-based multi-agent reinforcement learning method with difference reward assignment for training the routing agents in the constructed environment. For the agent credit assignment, we introduce difference reward for agents and adopt a counterfactual-inspired independent reward assignment mechanism inspired by \cite{foerster2018counterfactual} for achieving a high TE performance. Finally, the learned agents intelligently and timely generate the adaptive routing policies to control the forwarding behaviors of SDN switches in the highly dynamic environment. Through extensive experimental results and evaluations, we demonstrate that our proposed method has a superior performance compared with the previous TE solutions. \par

In a nutshell, the main contributions of this paper can be summarized as follows:\par
\begin{itemize}
\item  We propose a multi-agent reinforcement learning based approach CMRL to efficiently solve the TE problem in the dynamic hybrid SDN, especially when the network scale or network complexity increases. Specifically, an interactive environment is first constructed for offline agent training. Next, the routing agents are trained offline in the constructed environment for collaboratively learning the map between the traffic demands and routing policies of SDN switches. Finally, the trained agents are deployed to enable a timely and intelligent routing policy inference when traffic demands change or network failures happen.
\item We integrate difference reward assignment among agents into offline agent training in order to solve the multi-agent credit assignment problem. Specifically, each agent is given a reward that computes the estimated return for the current joint action to a baseline that marginalises out the agent's action and with the other agents' actions unchanged. The difference reward assignment encourages different agents to sacrifice for the better actions in the cooperative setting.

\item We conduct extensive experiments on real network topologies and traffic traces to evaluate the performance of CMRL. The experimental results demonstrate the superior performance of our proposed method CMRL in improving TE performance of the hybrid SDN when traffic demands change or network failures happen.

\end{itemize}

The rest of the paper is organized as follows. Section \ref{s2} presents the related work on TE solutions. Section \ref{s3} provides the problem definition. In Section \ref{s4}, we describe our proposed method CMRL, including the offline training part and online inference part. Section \ref{s5} shows the experimental results of different algorithms on real network topologies and traffic traces when traffic demands change or network failures happen. Finally, we make the conclusion and give possible future research direction in section \ref{s6}.\par

\section{Related Work}\label{s2}

In this section, we present the related work on the TE solutions in traditional distributed network and Software Defined Networks, respectively. \par

\subsection{TE Solutions in Traditional Distributed Network}\label{SDN}%

In the traditional distributed network, distributed routing protocols dominate and the traffic is constrained to route on the shortest paths between the source and destination according to link weight setting under the distributed routing protocols, such as Open Shortest Path First (OSPF) protocol \cite{moy1998ospf}, Intermediate System to Intermediate System (IS-IS) protocol \cite{oran1990rfc1142}. The link weight setting under the distributed routing protocols determines the available shortest paths for routing and many related works focus on optimizing distributed link weight setting using heuristic algorithms \cite{fortz2002traffic, ericsson2002genetic, wang2008overview} or machine learning approach based on gradient descent \cite{kodialam2022network}. However, for traditional distributed networks, flow routing lacks of flexibility with the shortest-path-routing constraint, which greatly limits the TE performance.\par

\subsection{TE Solutions in Software Defined Networks}\label{DRL} 

With the prevailing of SDN architecture, the routing gains more flexibility. The SDN controller can centrally control the forwarding behavior of SDN switches through dispatching flow entries and traffic can be routed on all available paths between a source-destination pair, regardless of shortest paths constraint. Microsoft \cite{hong2013achieving} and Google \cite{jain2013b4} have already built the SDN-enabled datacenter networks and have boosted the network utilization to near $100\%$ through flexible flow routing. However, fully deploying the SDN switches to replace the legacy router is a non-trivial task and encounters various challenges. To incrementally deploy the SDN switches into traditional network, Agarwal et al. \cite{agarwal2013traffic} propose a greedy approach to determine the placement of SDN switches. To improve the TE performance of the hybrid SDN, a Fully Polynomial Time Approximation Scheme is also introduced in \cite{agarwal2013traffic} to optimize the traffic splitting ratio at SDN switches. To further reduce the MLU of the hybrid SDN, Guo et al. \cite{guo2014traffic,guo2019sote} propose heuristic algorithms that jointly optimize the OSPF link weight setting and traffic splitting ratio at SDN switches under static and dynamic environments. To maximize the network throughput in hybrid SDN, Xu et al. \cite{xu2017incremental} introduce an approximation algorithm to optimize flow routing on $h$ available paths between a source-destination pair.\par

With the significant advances of reinforcement learning in various fields, more researchers begin to apply them in solving TE problems. Noting that the modern communication networks are becoming more complicated and highly dynamic, Xu et al. \cite{xu2018experience} and Chen et al.\cite{chen2020rl} propose novel experience-driven model-free Deep Reinforcement Learning (DRL) methods for solving TE problem. To mitigate the impact of network disturbance, Zhang et al.\cite{zhang2020cfr} introduce a reinforcement learning method CFR-RL for selecting and rerouting the critical flows to balance the network link utilization. To better optimize the link weight setting under OSPF protocol in a hybrid SRv6 network, Tian et al. \cite{tian2020traffic} propose a DRL-based method for learning the optimal OSPF link weight setting under the given traffic demands in a trail and error manner. \par

To efficiently react to dynamic environment and obtain the routing policy in an online manner, Guo et al. \cite{guo2021traffic} adopt a DRL method ROAR for learning the mapping between the traffic demands and routing polices. The trained agent can promptly infer the optimal routing policy when traffic demands change or network failures happen. To solve the TE problem in a multi-region scenario, Geng et al. \cite{geng2020multi,geng2021distributed} propose a distributed TE framework based on DRL for optimizing route selection under highly dynamic traffic. To optimize route selection of traffic flows under QoS requirements, Liu et al.\cite{liu2021drl} propose an online routing algorithm DRL-OR based on DRL for computing the optimal next hop at each router and further introduce safe learning mechanism to facilitate the online learning process.\par

Previous TE works in a hybrid SDN either leverage heuristic algorithms or single-agent reinforcement learning for optimizing network performance. As with the rapid increasing of network scales and network complexity, previous TE solutions suffer a performance degradation when traffic demands fluctuate or network failures happen because of high delay in computing routing policies. In this paper, we innovatively leverage the multi-agent reinforcement learning with a combination of difference reward assignment mechanism to improve the TE performance in a dynamic hybrid SDN environment.\par

\section{Problem Definition}\label{s3}

In our hybrid SDN network environment, the network topology is denoted as an undirected graph $G = (V,E)$. Here, $V$ represents the set of forwarding devices, which is composed of SDN switches $V_s$ and legacy routers $V_l$. $E$ refers to the set of links, where the capacity of each link $e \in E$ is represented as $C(e)$. Here, the link capacity refers to the maximum traffic volume that the link can accommodate. The Traffic Matrix (TM) is represented by $D$, which is a set of total traffic demands that need to be delivered from the source to the destination. The element $D$ ($u$, $v$) in TM $D$ represents the total traffic demand from source node $u$ to destination node $v$. The traffic routing in TE is to distribute the traffic demands onto the available candidate paths between the source and the destination to optimize maximum link utilization, network cost, or network throughput, etc. Specifically, at the legacy routers, the traffic is routed to the next hops on the shortest paths. At the SDN switches, the traffic can be flexibly routed on the multiple next hops to the destination. We refer to the common goal of TE, i.e., minimizing the Maximum Link Utilization (MLU) of a hybrid SDN without violating the link capacity constraints, as our TE optimization goal in this paper. Here, link utilization refers to the ratio of the link load to link capacity (bandwidth) and MLU indicates the largest link utilization of all links. Our approach can also be easily extended and adopted to optimize other TE goals in a hybrid SDN.\par

\section{Proposed Method CMRL}\label{s4}


%

In this section, we first provide an overview of the proposed method CMRL. Then, we exhibit the offline agents training phase of CMRL for learning the hidden patterns between the traffic demands and routing policy. Finally, with the trained agents deployed, the online routing inference phase of CMRL is elaborated on to exhibit the routing policy generation process when new traffic demands arrive. \par

\subsection{An overview of CMRL}\label{An overview of CTDE}

\begin{figure}[!htb]
	\begin{center}
		\includegraphics[width=7cm]{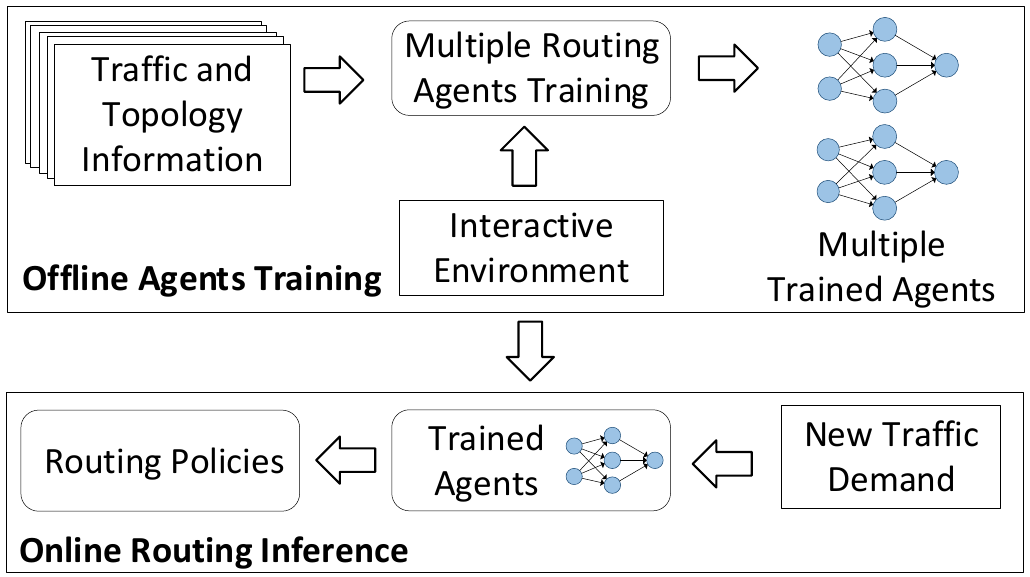}
		\caption{An overview of CMRL. }
		\label{overview}
	\end{center}
\vspace{-0.3cm}
\end{figure}

Our proposed method CMRL consists of two phases: offline agents training and online routing inference. An overview of CMRL is presented in Fig. \ref{overview}.\par

In the offline agents training, to train these agents, which are denoted by Deep Neural Networks (DNNs), an interactive environment should be first constructed for the interaction with the multiple agents. Then, given the traffic and topology information, multiple routing agents collaboratively learn the mapping between the traffic demands and routing policies through the interaction with the interactive environment. The well trained agents are deployed at SDN switches for online routing inference. \par

In the online routing inference, given the new arrival traffic demands and partial observed adjacent link utilization information, the distributedly deployed trained agents can infer the routing policy promptly with the trained DNNs. In the following, we will introduce the two phases in details.


\begin{figure*}[!htb]
	\begin{center}
		\includegraphics[width=15cm]{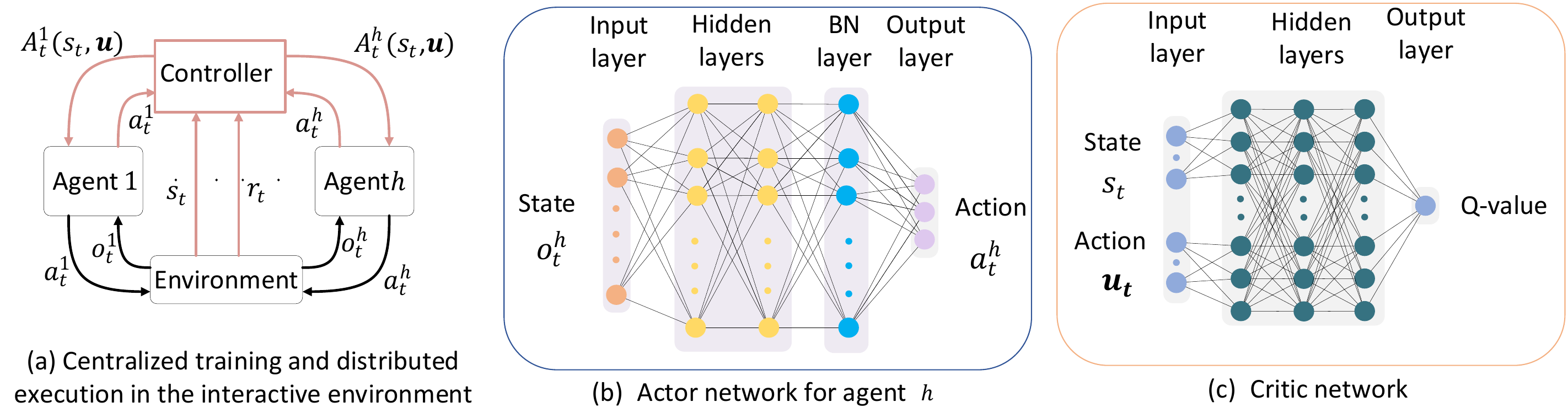}
		\caption{The architectures of multi-agent reinforcement learning, actor network and critic network. In (a), the red arrows and components are only required in centralized training. }
		\label{AC}
	\end{center}
\vspace{-0.5cm}
\end{figure*}

\subsection{Offline Agents Training}\label{offlinetraining}

To enable routing agents intelligently and efficiently learn the mapping between the traffic demands and routing policy, we adopt to Deep Deterministic Policy Gradient (DDPG) \cite{lillicrap2015continuous}, which is an off-policy, model-free, actor-critic DRL algorithm, for training the agents through the interaction with environment. Compared to other DRL algorithms, DDPG can handle the continuous high-dimension action space and state space, which is suitable for learning the continuous splitting ratios of traffic at SDN switches. In this part, we first elaborate on the interactive environment construction for training the multiple DRL agents offline. Then, given the interactive environment, we introduce the DDPG-based offline multi-agent training process.

\subsubsection{\textbf{Interactive Environment}}

In the offline training phase, the multiple agents collaboratively learn the routing policy through interacting with the environment. Therefore, the interactive environment should be constructed first. As shown in Fig. \ref{AC} (a), in each time step $t$, each agent $h$ has a partial observation of the network status $o_t^h$ and chooses corresponding action $a_t^h$ according to the partial observed environment state $o_t^h$. After agents taking the action, the environment provides a reward $r_t$ for the joint action $\boldsymbol {u_t}=\{a_t^h\}_{h=1}^H$ and the centralized controller receives the reward $r_t$ that indicates the performance of the joint action $\boldsymbol {u_t}$. Here, $H$ denotes the total number of agents. To make the agents learn the better action and achieve the global TE goal, we introduce the difference reward assignment for each agent. The difference reward assignment, denoted as $A_t^h(s_t,\boldsymbol u)$, is computed by the controller for specifying the contribution of each agent and participates in the actor network update. Then, the environment transits from $s_t$ to the new state $s_{t+1}$. The past experiences, with a form of $(\{o^h_0\}_{h=1}^H, \boldsymbol {u_t}, s_t, r_t, s_{t+1})$, are restored in the replay buffer for agent learning. An optimal policy $\{\pi_h^*\}$, which maps the partial observed network state to a probability distribution over actions, maximizes the expected cumulative discounted reward $E_{\{\pi_h^*\}}[\sum_{t}\gamma^{t}{r_t}]$. Here, the term $\gamma \in [0,1]$ is the discount factor which prevents an infinite sum of accumulated rewards. To be specific, the state, action and reward functions are designed as follows.\par


\textbf{State: }
The state is the input to the agents and should be carefully designed. The link utilization, which is determined by both the traffic demands and routing policy, can well reflect the network status. To better describe the network status and minimize the MLU in a hybrid SDN, we define the entire environment state $s_t$ at time $t$ based on the utilization of each link $s_t^e$, i.e., $s_t=\{s_t^e, e\in E\}$. The link utilization $s_t^e$ can be easily computed by the link load and the link capacity shown in Eq. (\ref{state}). \par

\begin{equation}\label{state}
	{s}_t^e = \frac{\sum\limits_{q \in V} f_{e, D_{t}}^{q}}{C(e)}	
\end{equation}
Here, $f_{e, D_{t}}^{q}$ is the splitting traffic on link $e$ destined to $q$ under TM $D_t$. $\sum\limits_{q \in V} f_{e, D_{t}}^{q}$ calculates the total traffic volume on link $e$ under TM $D_t$.\par

In a practical SDN-TE system, each SDN switch can obtain the utilization of all links through the information flooded regularly by other switches. However, leveraging the utilization information of all links leads to the slow convergence of the method. To improve the convergence speed of CMRL and better guide the training of DNNs in agents, we exploit the partial observation of each SDN switch instead of the entire link utilization in offline agents training. Here, the partial observation of the SDN switch refers to the adjacent link utilization of the switch and we use $o^h_t$ to denote the partial observation of the SDN switch with the deployed agent $h$ at time $t$, $h\in[1,H]$.

\textbf{Action:}
The action is the output of the agent which indicates the routing policy of the deployed SDN switch. The action $a_t^h$ of agent $h$ at time $t$ is a vector that consists of traffic-splitting ratios $\chi_{e, D_{t}}^{h}$ distributed to the adjacent links $e$ of SDN switch $h$ under TM $D_t$ as shown in Eq. (\ref{action}). It should be noted that the sum of the splitting ratios of the traffic on all the adjacent links of each SDN node should be equal to $1$. Therefore, we adopt to a normalization of the output to the actor networks for guaranteeing this constraint. The joint actions $\boldsymbol u_t$ of the multiple agents are defined as $\boldsymbol {u_t}=\{a_t^h\}_{h=1}^H $. \par

\begin{equation}
	a_t^h=\{\chi_{e, D_{t}}^{h}\}_{p_e=h} 	\label{action}
\end{equation}
$p_e$ denotes the head of the link $e$.

\textbf{Reward: } The agents work collaboratively for achieving a global TE goal, which is to minimize the MLU of the hybrid network. Therefore, we design a reward function which can reflect the global TE performance of the hybrid SDN. The reward function related to MLU is shown in Eq. (\ref{eq:reward2}):

\begin{equation}
r_t=
\begin{cases}
-e^{2(\frac{1}{\alpha}-1)} & \mbox{if $\alpha$ $\textgreater$ $1$} \\
0 & \mbox{if $\alpha$ = $1$}\\
e^{2(\alpha-1)} & \mbox{if $\alpha$ $\textless$ $1$}
\end{cases}
\label{eq:reward2}
\end{equation}

\begin{equation}
\alpha=\frac{U_{i,0}^{max}}{U_{i,t}^{max}}
\label{eq:reward1}
\end{equation}
Here, $U_{i,0}^{max}$ and $U_{i,t}^{max}$  denotes the MLU of the network under the TM $D_i$ at time $0$ and $t$, respectively. We treat $U_{i,0}^{max}$ as a baseline in minimizing MLU, which is computed by routing all the flows to the next hops on the shortest paths according to the OSPF protocol. The $\alpha$ value, which is computed by Eq. (\ref{eq:reward1}), shows the improvement ratio of the network performance when the new routing policies are deployed at time $t$. As shown in Eq. (\ref{eq:reward2}), a higher positive reward $r_t$ is given if the joint action taken by multiple agents at time $t$ generates a lower MLU which improves network performance. Otherwise, a lower negative reward is given if the action taken generates a higher MLU that degrades the network performance.

\subsubsection{\textbf{Offline Training}}


Given the interactive environment, multiple routing agents are trained offline with DDPG for learning the mapping between the traffic demands and routing policy. Each agent maintains actor networks for approximating the mapping between the input network status to routing policy outputs. There is also a critic network for evaluating the actions taken by the agents and helping update the parameters of actor networks. Specifically, the actor network of each agent outputs an action $a_t^h$ based on its partially observed state $o_t^h$, and the critic network generates a reward $r_t$ given the joint action $\boldsymbol {u_t}$ and partially observed state $o_t^h$ of each agent $h$. Both the actor network and the critic network have two sub-networks, the online network and the target network, respectively and the structures of the two sub-networks are the same. \par

As shown in Fig. \ref{AC}(b), the actor network consists of input layer, hidden layer, Batch Normalization (BN) layer and output layer. The two hidden layers are fully-connected layers and each layer contains 1024 neurons. The activation function is ReLU. The BN layer is added to the actor network in order to speed up the training and convergence of the neural network as well as to control the gradient exploding and prevent the gradient vanishing. The output layer refers to Softmax as its activation function to ensure that the sum of the splitting ratios for a demand equals $1$. For evaluating the joint action taken by agents, a critic network is kept at the centralized controller to generate a Q-value for the joint action. As shown in Fig. \ref{AC}(c), the critic network consists of an input layer, three hidden layers and an output layer. Both three fully-connected hidden layers have 1024 neurons and the activation function is also ReLU. The historical data of actor network and critic network are stored in the replay buffer $B$ for training.\par


Algorithm \ref{alg1} presents the offline training phase of multiple agents. The inputs in Algorithm \ref{alg1} contain the hybrid SDN environment $G$, the historical TMs ${\emph{D}}$, the set of available links $\emph{L}$ for each traffic demand, and the number of agents $\emph{H}$. \par
\begin{breakablealgorithm}
	\setstretch{1.1} 
    \label{alg1}
	\caption{Offline agents training}
	\begin{algorithmic}[1]
	\Require $G=(V,E)$, $\emph{D}$, $\emph{L}$, $\emph{H}$
	\Ensure $\boldsymbol \mu = {\{\mu_h({\boldsymbol{o_h}}|\theta_h^\mu)\}}_{h=1}^{H}$
	\State Initialize action set$ $ $\boldsymbol {u}=\{a_h\}_{h=1}^H$;\;\label{r1}
	\State Initialize $H$ online actor networks   ${\{\mu_h({\boldsymbol{o_h}}|\theta_h^\mu)\}}_{h=1}^{H}$ and an online critic network ${Q(s,\boldsymbol u|\theta^Q)}$ with random 	$\{\theta_h^\mu\}_{h=1}^{H}$ and $ \theta^Q$;\;\label{r2}

	\State Initialize $H$ target actor networks  ${\{\mu'_h}({\boldsymbol {o_h}}|\theta_h^{\mu'})\}_{h=1}^{H}$  and a target critic network ${Q'}(s,\boldsymbol u|\theta^{Q'})$ with 		$\{\theta_h^{\mu'}\}_{h=1}^{H}\leftarrow\{\theta_h^\mu\}_{h=1}^{H}$, $\theta^{Q'}\leftarrow\theta^Q$;\;\label{r3}
	\State Initialize replay buffer $B$ ;\;\label{r4}
	\For{$D_{i}$ in $\textbf{\emph{D}}$}
	\State Initialize OU process ${\boldsymbol N}=\{\mathcal{N}^h\}_{h=1}^{H}$;\;\label{r5}

	\State $\epsilon = 1.0$;\;\label{r6}
	\label{code:endinit}
	\For{episode $n=1\cdot\cdot\cdot N$}
	\label{code:beginiter}
	\label{code:begininitstate}
	\State $F_{i,0}=$get\_ospf\_flows$(G, D_i)$;\;\label{r7}
    \State $ (\{o^h_0\}_{h=1}^H, s_0, U_{i,0}^{max})=$get\_state$(F_{i,0})$;\;\label{r8}
	\label{code:endinitstate}
	\For{step $t=0 \cdot\cdot\cdot {T-1}$}
	\label{code:begint}
	\For{agent $h=1 \cdot\cdot\cdot H$}
	\label{code:begint}\label{r9}
	\State $a^h_t$=$\mu_h(o^h_t|\theta_h^\mu)+\epsilon \mathcal{N}^{h}_{(n-1)T+t},\epsilon\leftarrow\eta\epsilon$;\;
	\label{code:begintransition}
	\EndFor\label{r10}
	\State $\boldsymbol{ u_t} = \{ a^h_t\}_{h=1}^H $;\;
	\State $\textbf{\emph{P}}=$ get\_policy$({\boldsymbol {u_t}, \emph{L}})$;\;\label{r11}

	\State $F_{i,t+1}=$ get\_flow$s(G, D_i, \textbf{\emph{P}})$;\;
    \State $(\{o^h_{t+1}\}_{h=1}^H,s_{t+1}, U^{max}_{i,t+1})$ =get\_state$(F_{i,t+1})$;\;\label{r15}
	\State $r_{t}=$get\_reward$(U^{max}_{i,t+1},U^{max}_{i,0})$;\;\label{r16}
	\If{$t =T$}\label{r12}
	\State $done_t$= 1;\;
	\Else
	\State $done_t$= 0;\;
	\EndIf\label{r13}
	\State $B$.store$(\{o^h_t\}_{h=1}^H,\boldsymbol {u_t}, s_t,r_t, s_{t+1}, done_t)$;\;\label{r14}
	\label{code:endtransition}
	\State minibatch $B'=B$.sample$(M)$;\;\label{r17}
	\label{code:begintrain}

	\label{code:begintrain}
	\For{$(\{o^h_j\}_{h=1}^H, \boldsymbol {u_{j}},s_{j}, r_{j}, s_{j+1}, done_j) \in B'$}
	\State $y_j = r_j + \gamma(1 - done_j)Q'(s_{j+1},\{\mu_h'(o_{j+1}|\theta^{\mu'}_h)\} _{h=1}^H|$  $\theta^{Q'})$;\;
	\EndFor
	\State Update online critic and actor with Eq. \eqref{eq:update4}-\eqref{eq:update6};\;
	\State Update target critic and actor with  Eq. \eqref{eq:update7}\eqref{eq:update8};\;\label{r18}
	\EndFor
	\label{code:endt}
	\EndFor
	\label{code:enditer}
	\EndFor
	\State \Return $\boldsymbol  \mu$ ;\;
    \end{algorithmic}
\end{breakablealgorithm}

In Algorithm 1, we start with initialization (lines \ref{r1}-\ref{r4}). The online actor networks ${\{\mu_h({\boldsymbol{o_h}}|\theta_h^\mu)\}}_{h=1}^{H}$ and online critic network ${Q(s,\boldsymbol u|\theta^Q)}$ are initialized using random $\{\theta_h^\mu\}_{h=1}^{H}$ and $\theta^Q$, respectively (line \ref{r1}-\ref{r2}). The target actor networks ${\{\mu'_h}({\boldsymbol {o_h}}|\theta_h^{\mu'})\}_{h=1}^{H}$  and target critic network ${Q'}(s,\boldsymbol u|\theta^{Q'})$ are initialized with the same parameters as the online networks (line \ref{r3}). The replay buffer $B$ is initialized as a circular array with a fixed size for agent learning (line \ref{r4}).\par

Then, we begin to train the multiple agents under the historical TM set $\emph{\textbf{D}}$. To better explore the action space, the Ornstein-Uhlenbeck ($OU$) process $\boldsymbol N$ is initialized  for action exploration and parameter $\epsilon$ is initialized to 1.0, which is used to balance action exploration and exploitation (line \ref{r5}-\ref{r6}). At the beginning of each episode, through function $get\_ospf\_flow$ and $get\_state$, we calculate the initial state $s_0$ (line \ref{r7}-\ref{r8}). Here, $get\_ospf\_flow$ function derives the link load distribution $F_{i,0}$ with flows are routed on the shortest paths according to OSPF protocol and $get\_state$ function calculates the state $s_0$ according to Eq. (\ref{state}) given the traffic distribution $F_{i,0}$.\par

For each step $t$, the agent $h$ computes its own action $a_t^h$ based on the actor network $\mu(o^h_t|\theta_h^\mu)$ with a local partial observation $o^h_t$ and a discount OU noise $\epsilon\mathcal{N}^{h}_{(n-1)T+t}$ (line \ref{r9}-\ref{r10}). With the joint action $\boldsymbol{u_t} = \{ a^h_t\}_{h=1}^H$, we can derive the routing policy $P$ through the $get\_policy$ function and obtain the MLU of the hybrid SDN through $get\_state$ function (line \ref{r11}-\ref{r15}). Then, reward $r_t$ can be computed by Eq. (\ref{eq:reward2}) in $get\_reward$ function (line \ref{r16}). Here, $done_t$ implies whether the agent can get more reward in the rest of the steps (line \ref{r12}-\ref{r13}). Afterwards, transitions in the form of $(\{o^h_t\}_{h=1}^H, \boldsymbol {u_t}, s_t, r_t, s_{t+1}, done_t)$ are deposited in the replay buffer $B$ (line \ref{r14}). Finally, the critic and actor networks are updated on the randomly sampled minibatch $B'$ with $M$ transitions from the replay buffer $B$ (line \ref{r17}-\ref{r18} ). \par

Specifically, for the online critic network, the parameter $\theta^Q$ is updated by minimizing the loss between the cumulative reward $y_m$ and the estimated reward $Q(s_m, \boldsymbol {u_m}|\theta^Q)$ shown as follows.

\begin{equation}
L_Q(\theta^Q) = \frac{1}{M}\sum_{m=1}^M(y_m-Q(s_m, \boldsymbol{u_m}|\theta^Q))^2
\label{eq:update4}
\end{equation}

To encourage multiple agents to explore the good action, we introduce the idea of difference reward assignment into agent training. In actor-critic approaches, the reward of each agent is denoted as $A^h(s_t,\boldsymbol u)$ shown in Eq. (\ref{eq:update5}). In Eq. (\ref{eq:update5}), the first term on the right side $Q(s_t, \boldsymbol u)$ estimates the Q-value for the joint action $\boldsymbol u$ taken on the state of agent $h$. The advantage function $A^h(s_t,\boldsymbol u)$ is obtained by comparing the Q-value of the current action $a_t^h$ to a counterfactual baseline that marginalizes $a_t^h$ while keeping the actions of other agents $u^{-h}$ fixed.


\begin{equation}
A^{h}(s_t,\boldsymbol u) = Q(s_t, \boldsymbol u) -{\int Q(s_t,\boldsymbol {u'}=(\boldsymbol {u_t}, \boldsymbol {u_{t-1}^{-h}} )} ) d{\boldsymbol{u'}}
\label{eq:update5}
\end{equation}
Because the second integral term on the right is difficult to calculate, we adopt to Monte Carlo method \cite{shapiro2003monte} for sampling different actions in action space and calculate the Q value through the critic network.

Meanwhile, the parameter ${\theta^\mu_h}$ of the online actor network for agent $h$ is updated by the sampled policy gradient as follows.
\begin{equation}
\nabla J_\mu({\theta^\mu_h})\approx\frac{1}{M}\sum_{j=1}^M\ 	{A^h}(s_j^h, \boldsymbol u)\nabla_{\theta^\mu_h}{\mu}_h(o_m|\theta^\mu_h)
\label{eq:update6}
\end{equation}

For the target actor network and critic network, the parameters $\theta^{Q'}$ and ${\theta^{\mu'}_h}$ will be softly updated as follows.
\begin{equation}
\theta^{Q'} \leftarrow \tau \theta^{Q} + (1 - \tau)\theta^{Q'}
\label{eq:update7}
\end{equation}

\begin{equation}
\theta^{\mu'}_h \leftarrow \tau \theta^{\mu}_h + (1 - \tau)\theta^{\mu'}_h
\label{eq:update8}
\end{equation}

When the offline training process terminates, the outputs of Algorithm \ref{alg1} are the trained actor networks $\boldsymbol \mu$, which are the DNNs that have well learnt the mapping between the traffic demands and the routing policy.

\subsection{Online Routing Inference}
When offline agents training finishes, we deploy the trained routing agents on the SDN switches. When new traffic demands arrive, each agent generates the corresponding routing policy $\emph{P}$ in an online manner through the inference from the trained actor network as shown in Algorithm \ref{fatc}. The inputs to the algorithm are the network topology $G$, the set of trained actor networks $\bm{\mu}$, the set of available links $\emph{L}$, the current TM $D_i$, and the number of agents $H$. The learnt agents can derive the routing policy $\emph{P}$ based on the input information in $T$ steps.\par

\begin{algorithm}[htb]
\begin{algorithmic}[1]
 \Require $G=(V,E)$, $\bm{\mu}$, $\emph{L}$, $D_i$, $H$
  \Ensure $\emph{P}$, $U^{max}_{T-1}$
        \State $F_{i,0}=$get\_ospf\_flows$(G, D_i)$\\ $ (s_0, U_{i,0}^{max})=$get\_state$(F_{i,0})$
        \For{step $t=0$ to $T-1$}\label{e15}
	        \State $\boldsymbol{ u_t} = \{\mu_h(o^h_t|\theta_h^\mu)\}_{h=1}^H $;\;
            \State $\emph{P}=$ get\_policy$(\boldsymbol{ u_t}, \emph{L})$;\;
            \State $F_{i,t+1}=$ get\_flows$(G, D_i, \emph{P})$  ;\;
            \State $(s_{t+1}, U^{max}_{i,t+1}) =$ get\_state$(F_{i,t+1})$;\;
            \State $r_{t}=$ get\_reward$(U^{max}_{i,t+1}, U^{max}_{i,0})$;\;
        \EndFor \label{e16}\\
        \Return $\emph{P}, U^{max}_{T-1}$
  \caption{Online routing inference}\label{fatc}
  \end{algorithmic}
  \end{algorithm}

\section{Evaluation}\label{s5}
To demonstrate the superior performance of CMRL, we conduct extensive experiments on different network topologies and traffic datasets. In this section, we first introduce the environmental setup, including the dataset and baseline methods in section \ref{setting}. Then, we present the experimental results and analysis of various methods on TE performance under different traffic demands and network failure scenarios in section \ref{results}.    \par

\subsection{Experimental Setup}\label{setting}
The simulation experiments are executed on a workstation with eight Intel cores of 2.4GHz, a RAM of 256 GB and a NVIDIA GeForce RTX 3090 GPU. Our method is implemented on tensorflow. During the offline learning, the size of minibatch $M$ and the discount factor $\gamma$ are set to 32 and 0.9. The size of replay buffer $B$ and the term $\tau$ are set to 8000 and 0.001. The learning rates of the online actor and critic nets are set at $1\times 10^{-3}$ and $2\times 10^{-3}$. In addition, we set the number of episodes $N$ to 160. \par

\subsubsection{Dataset}
The experimental evaluation is carried out on three different network topologies: Abilene (12 nodes, 30 links), CERNET (14 nodes, 32 links) and G\'{E}ANT (23 nodes, 74 links), which are the research and education networks of America, China and European, respectively. The traffic demands datasets of Abilene and CERNET are provided by TOTEM \cite{balon2019traffic} and Zhang \cite{zhang2014cte}, which are measured every 5 minutes. The traffic demands dataset of G\'{E}ANT is provided by Uhlig \cite{uhlig2006providing}, which are measured every 15 minutes. In our experiments, $1024$ $(80\%)$ TMs are used for offline training and $256$ $(20\%)$ TMs are used for online inference.\par

%
%
%

\subsubsection{Baseline}
To exhibit the superiority of the proposed CMRL method, we conduct the comparative experiments with the following methods.\par
\begin{itemize}
	\item{\textbf{OSPF \cite{OSPF}:}} This method routes the network traffic according to the OSPF protocol. The traffic is routed on
	the shortest paths between source and destination node pair.
	\item{\textbf{ROAR \cite{guo2021traffic}:}} This method is a single-agent reinforcement learning approach that trains an intelligent routing agent with DDPG for minimizing the MLU of the hybrid SDN.
    \item{\textbf{MARL \cite{geng2020multi}:}} This method adopts to a general multi-agent reinforcement learning solution that learns the routing policy of each SDN switch without difference reward assignment for multiple agents.
	
\end{itemize}\par

\subsection{Experimental Results}\label{results}

\subsubsection{Parameter analysis}
\begin{figure}[!htb]
\vspace{-0.4cm}
	\begin{center}
		\includegraphics[width=6.6cm]{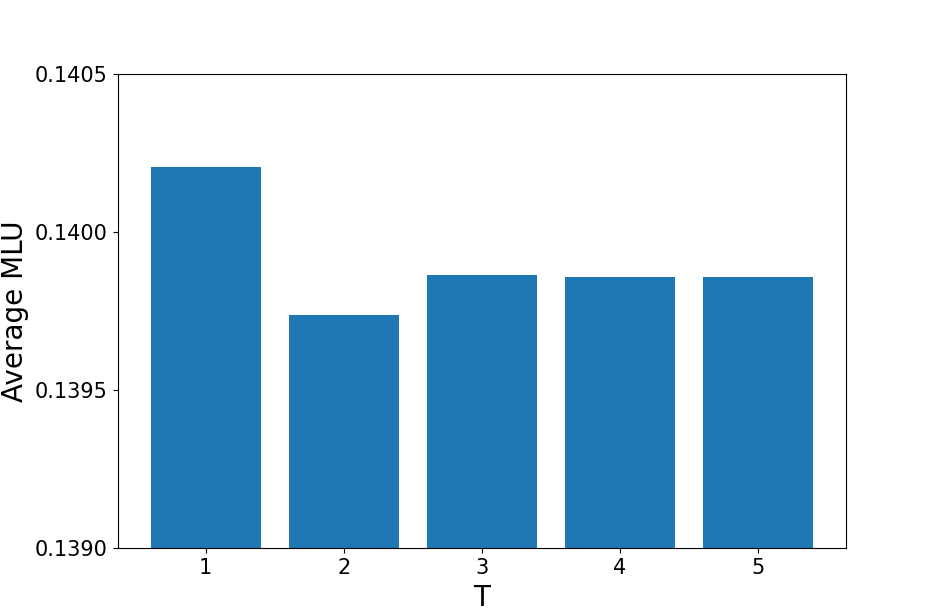}
		\caption{The average MLU under different values of $T$. }
		\label{step}
	\end{center}
\vspace{-0.6cm}
\end{figure}

The number of iteration times $T$ is an important parameter in online routing inference. We conduct experiments to evaluate the average MLU under different $T$ values as shown in Fig. \ref{step}. As shown in Fig. \ref{step}, we can observe that with the increasing of $T$, the average MLU first decreases, then increases and finally becomes flat. When $T=2$, the average MLU is the minimum. The reason is larger $T$ influences trained agent for finding the optimal routing policy. We set $T$ to $2$ in the following experiments.

\subsubsection{Convergence analysis}
\begin{figure}[!htb]
\vspace{0.2cm}
	\begin{center}
		\includegraphics[width=7cm]{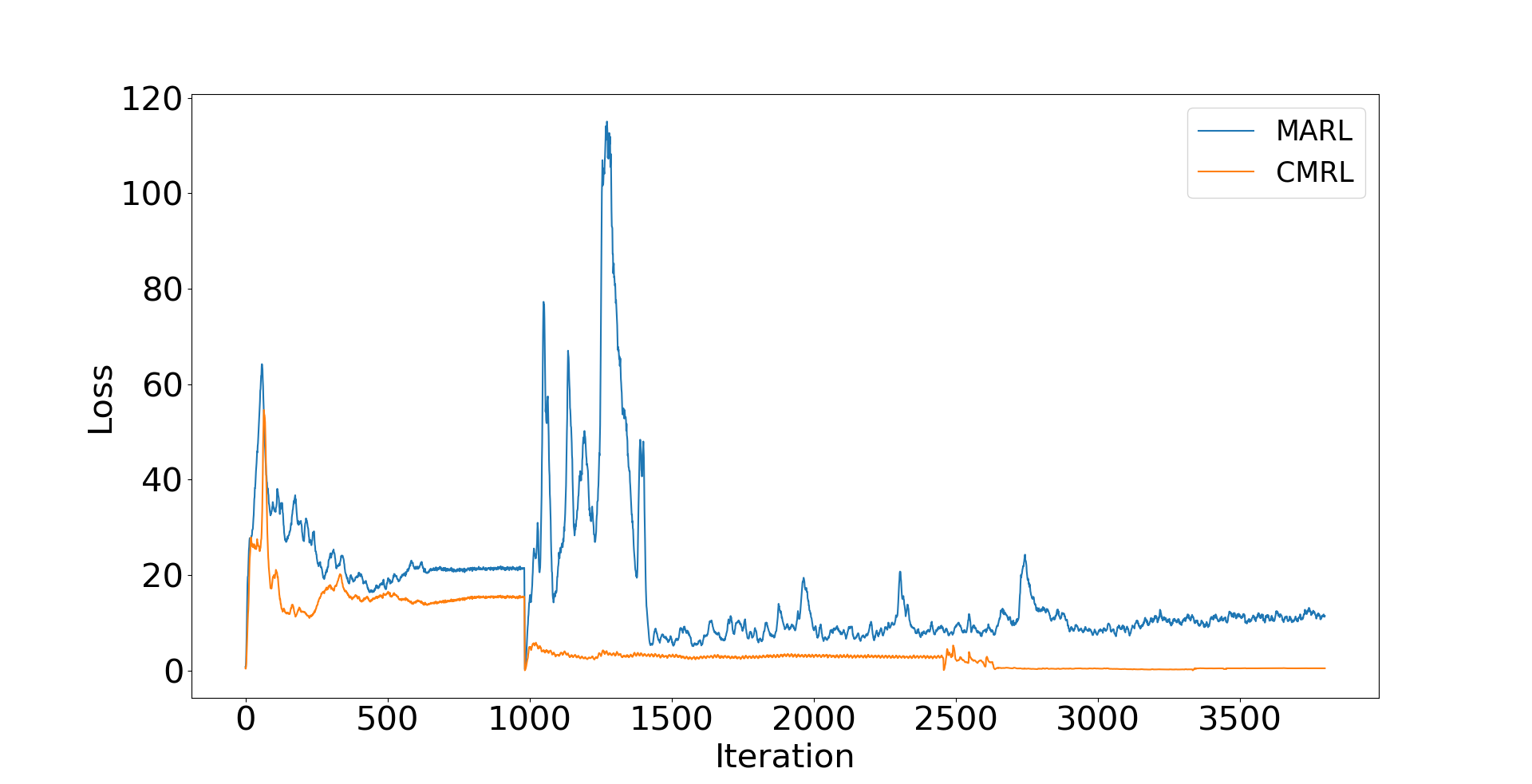}
		\caption{The training loss varies with the increasing of iteration times. }
		\label{los}
	\end{center}
\vspace{0.3cm}
\end{figure}

To demonstrate our proposed method CMRL converges, we plot the curve of training loss varies with the increasing of training episode in Fig. \ref{los}. As shown in Fig. \ref{los}, the training loss curves fluctuate with the increasing of training episodes. For the loss curve of CMRL, when the iteration time is smaller than $1000$, the loss value of CMRL fluctuates violently. When the iteration time is larger than $1000$, the loss curve becomes relatively flat and stays at a low value. The CMRL method converges when the iteration time reaches 1000. For the loss curve of MARL, the loss value fluctuates drastically at first. When the iteration time is larger than $3000$, the loss curve becomes relatively flat and stays at a low value. The MARL method converges when the iteration time reaches $3000$. The experiment demonstrates that our proposed method CMRL converges faster than the general MARL method.

\begin{figure*}[!htb]
\centering
\subfloat[0.2]{
\includegraphics[width=0.31\textwidth]{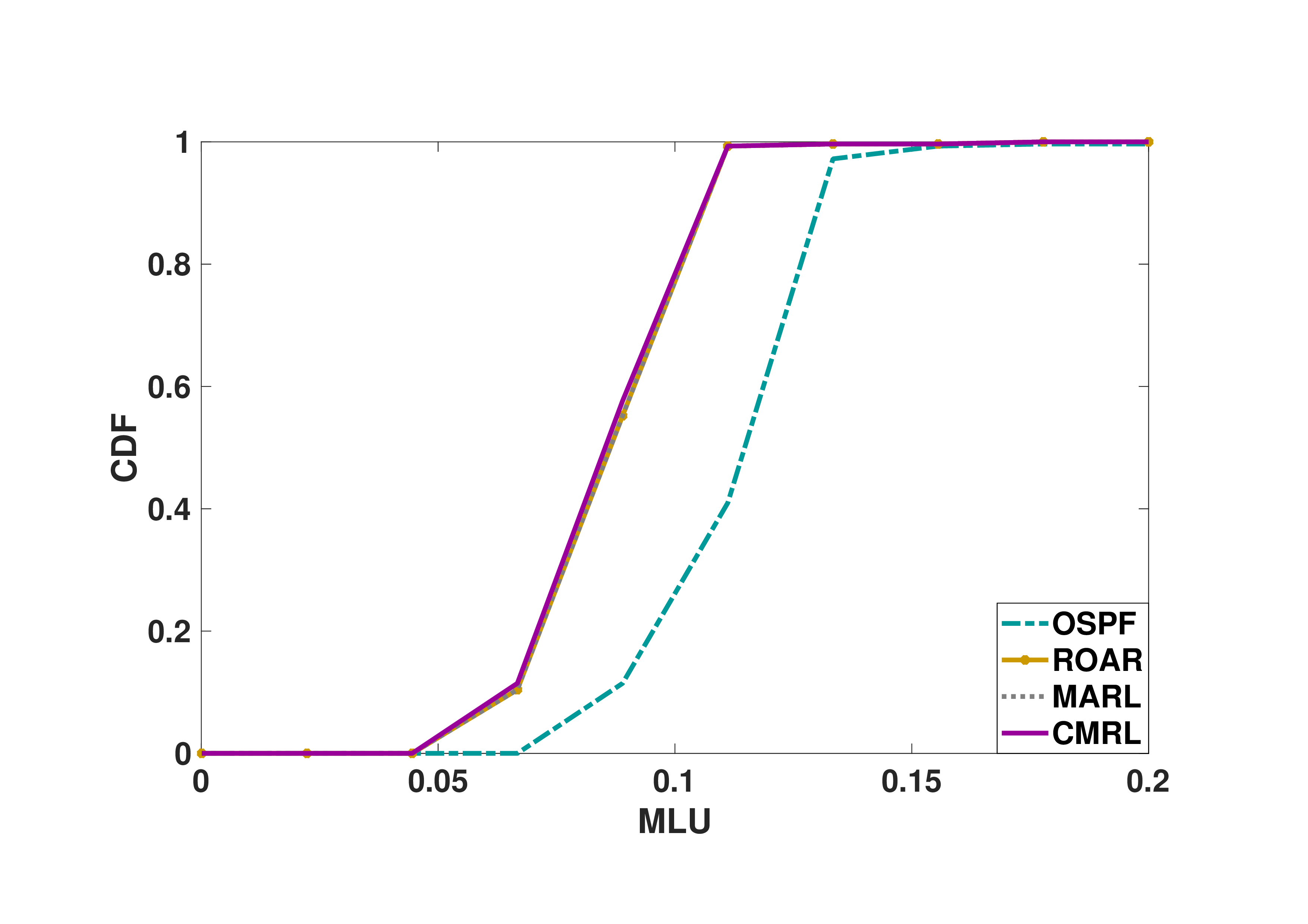}
}
\subfloat[0.3]{
\includegraphics[width=0.3\textwidth]{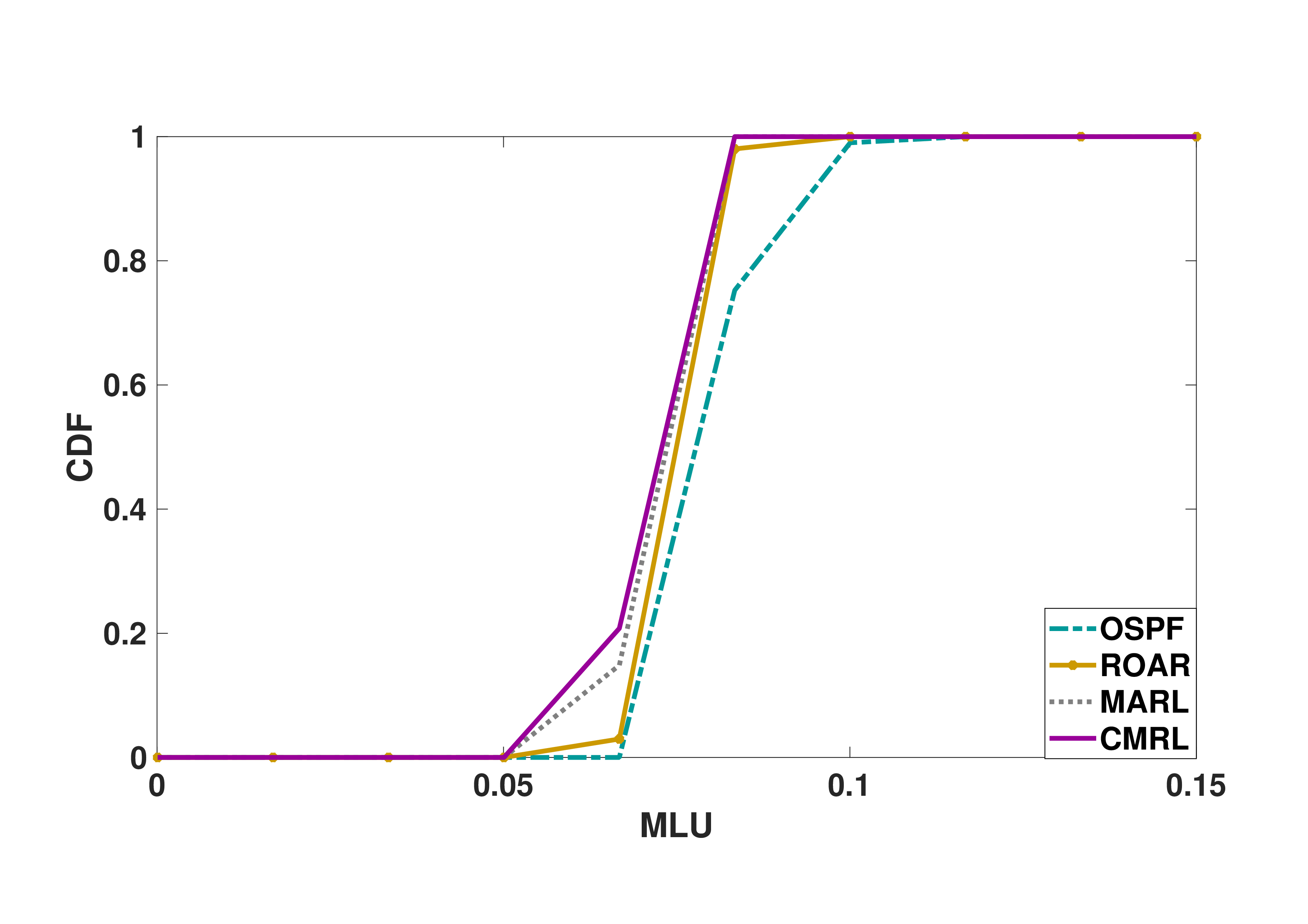}
}
\subfloat[0.4]{
\includegraphics[width=0.3\textwidth]{figure/12node3renew.png}
}
\vspace{-0.1cm} 
\caption{The CDF curves of MLU under different SDN deployment ratios in Abilene.}
\label{ab}
\end{figure*}

\begin{figure*}[!htb]
\vspace{-0.7cm} 
\centering
\subfloat[0.2]{
\includegraphics[width=0.3\textwidth]{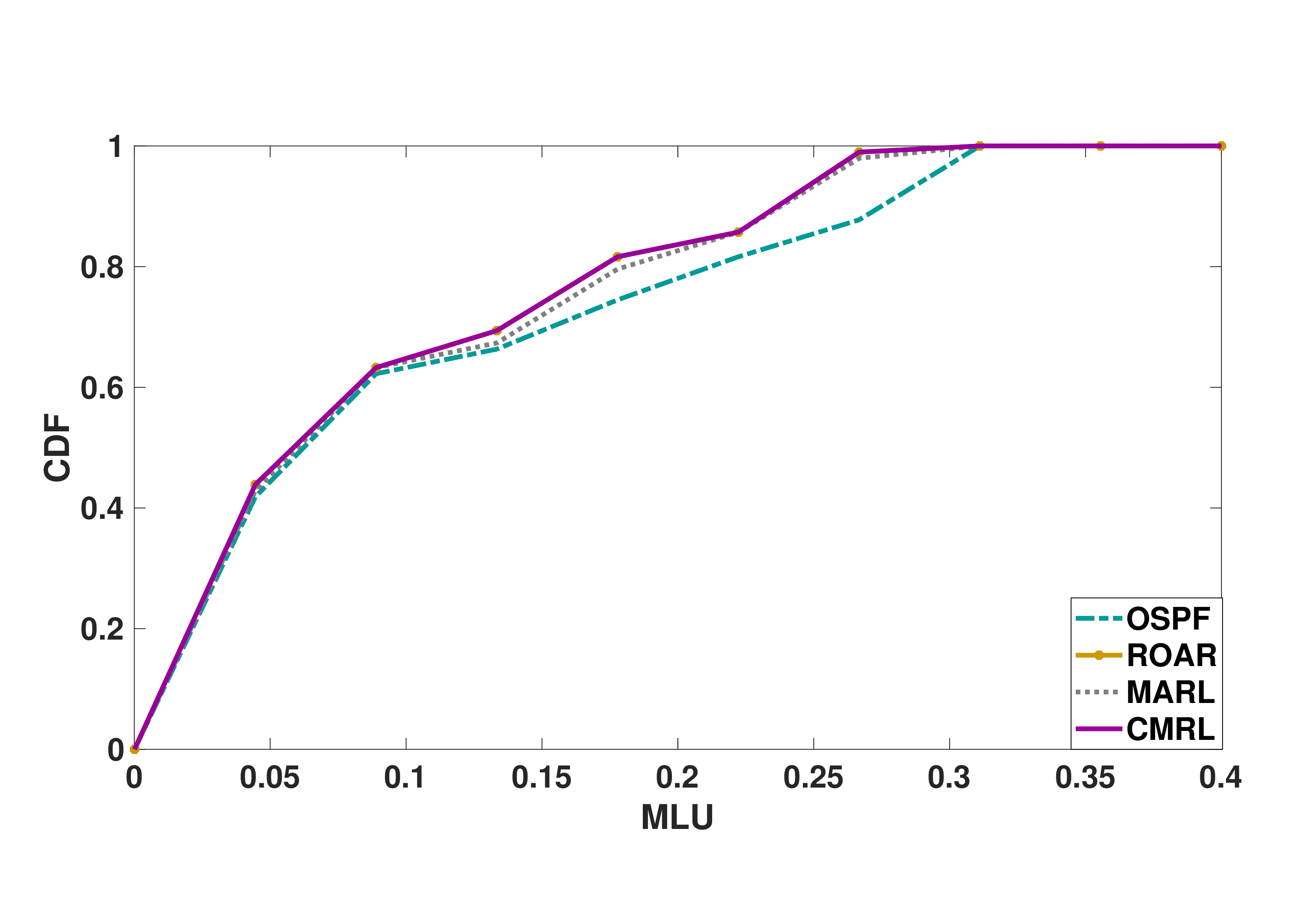}
}
\subfloat[0.3]{
\includegraphics[width=0.3\textwidth]{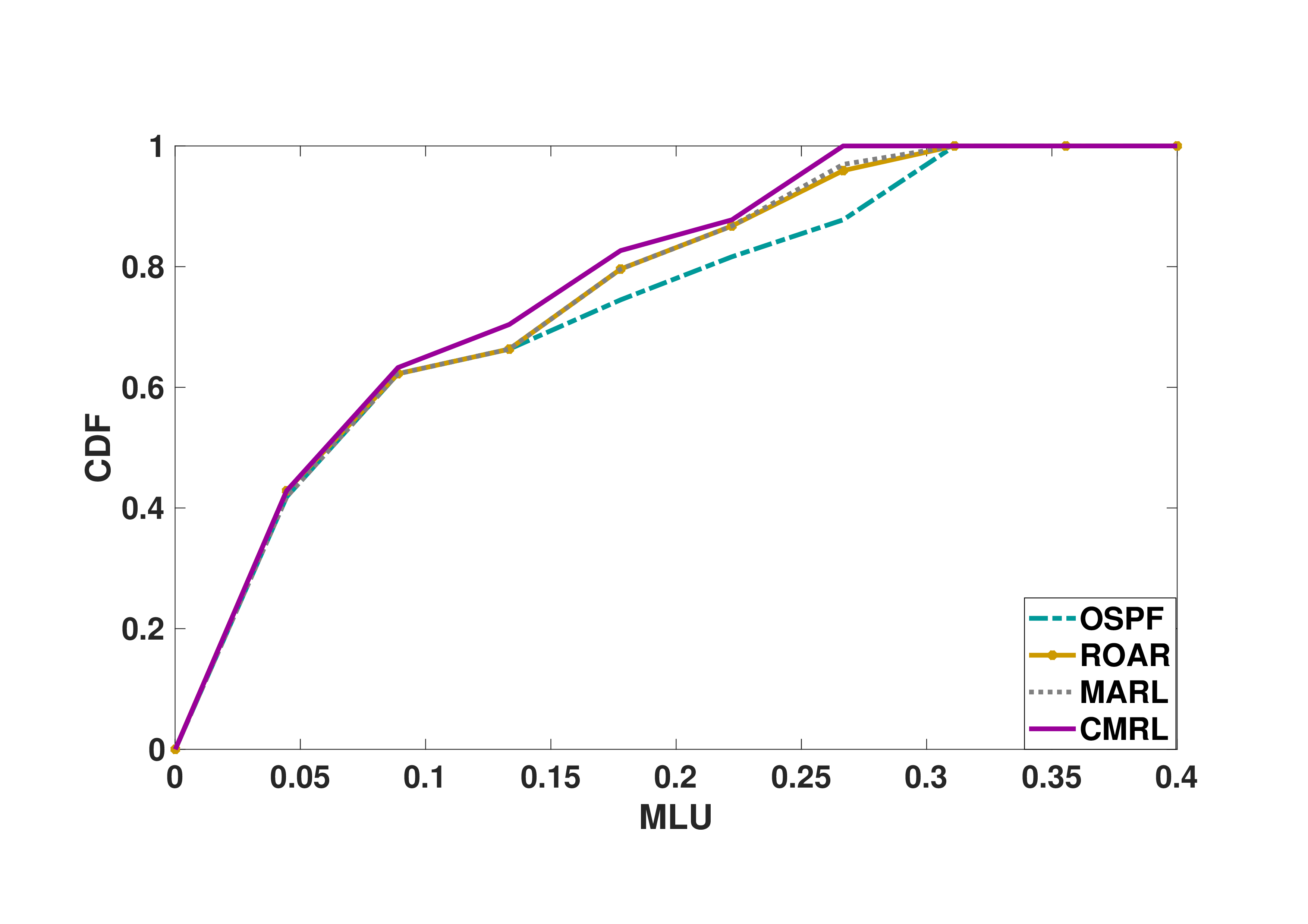}
}
\subfloat[0.4]{
\includegraphics[width=0.3\textwidth]{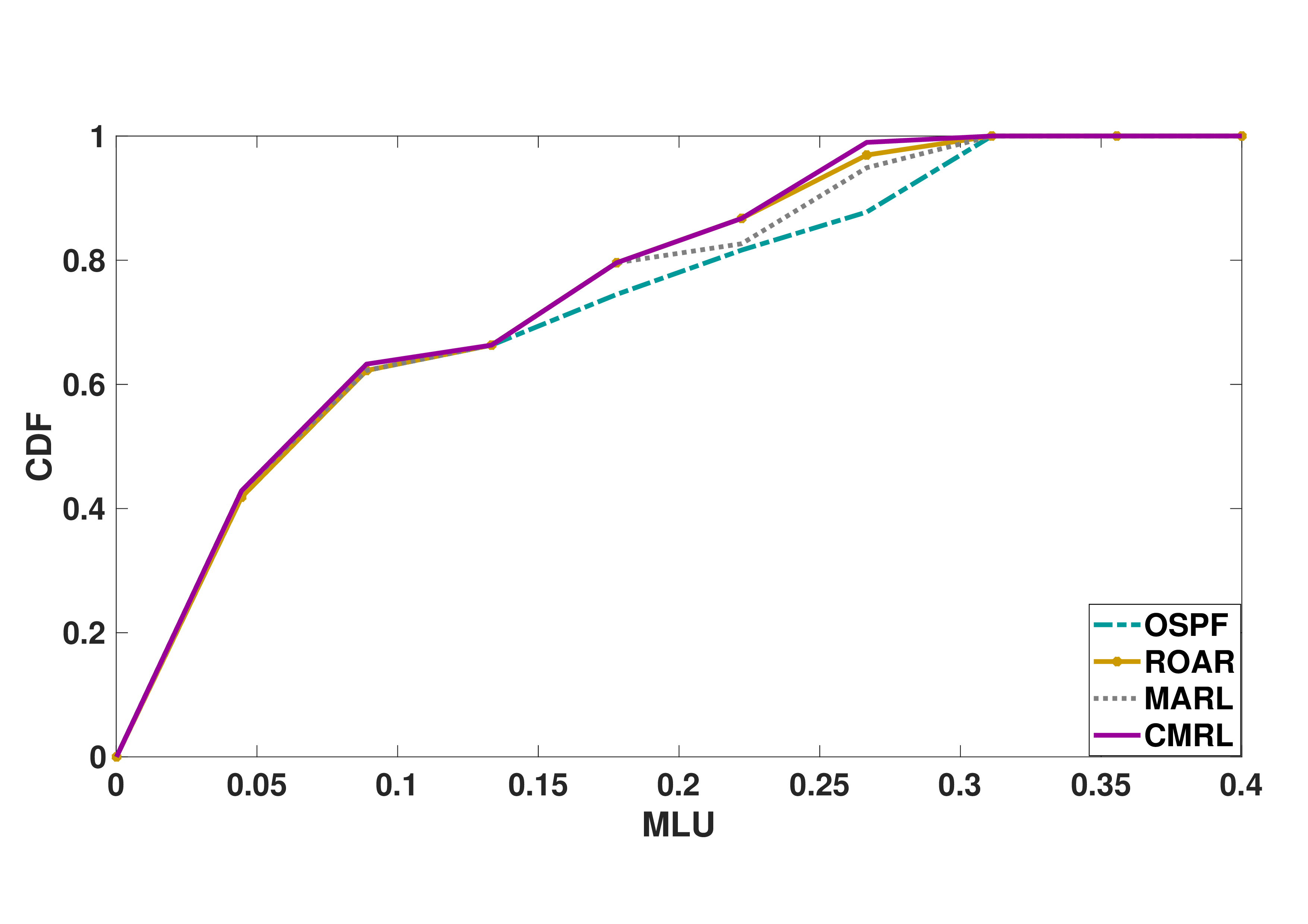}
}
\vspace{-0.2cm} 
\caption{The CDF curves of MLU under different SDN deployment ratios in CERNET.}
\label{cernet}
\end{figure*}

\begin{figure*}[!htb]
\vspace{-0.75cm} 
\centering
\subfloat[0.2]{
\includegraphics[width=0.3\textwidth]{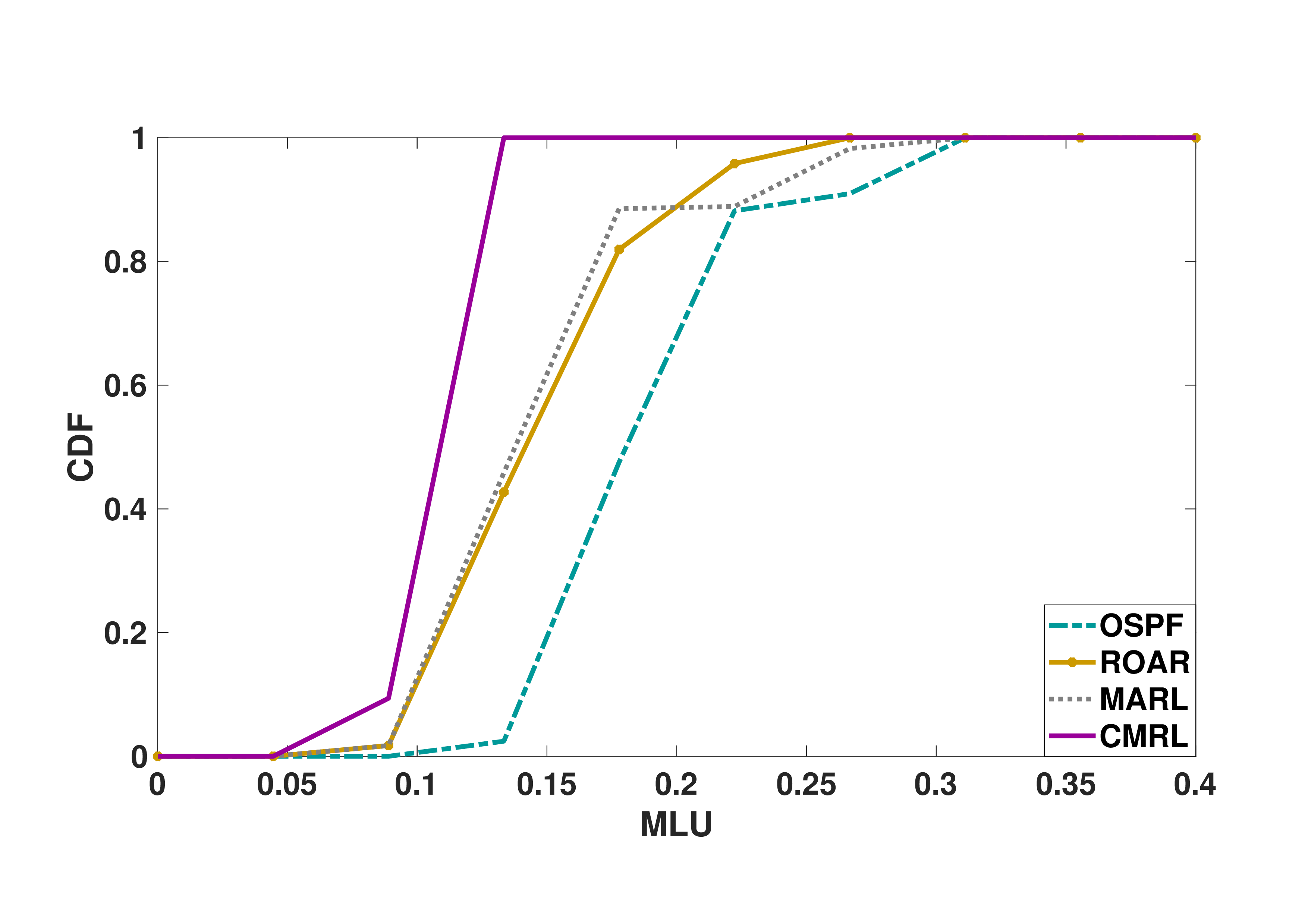}
}
\subfloat[0.3]{
\includegraphics[width=0.3\textwidth]{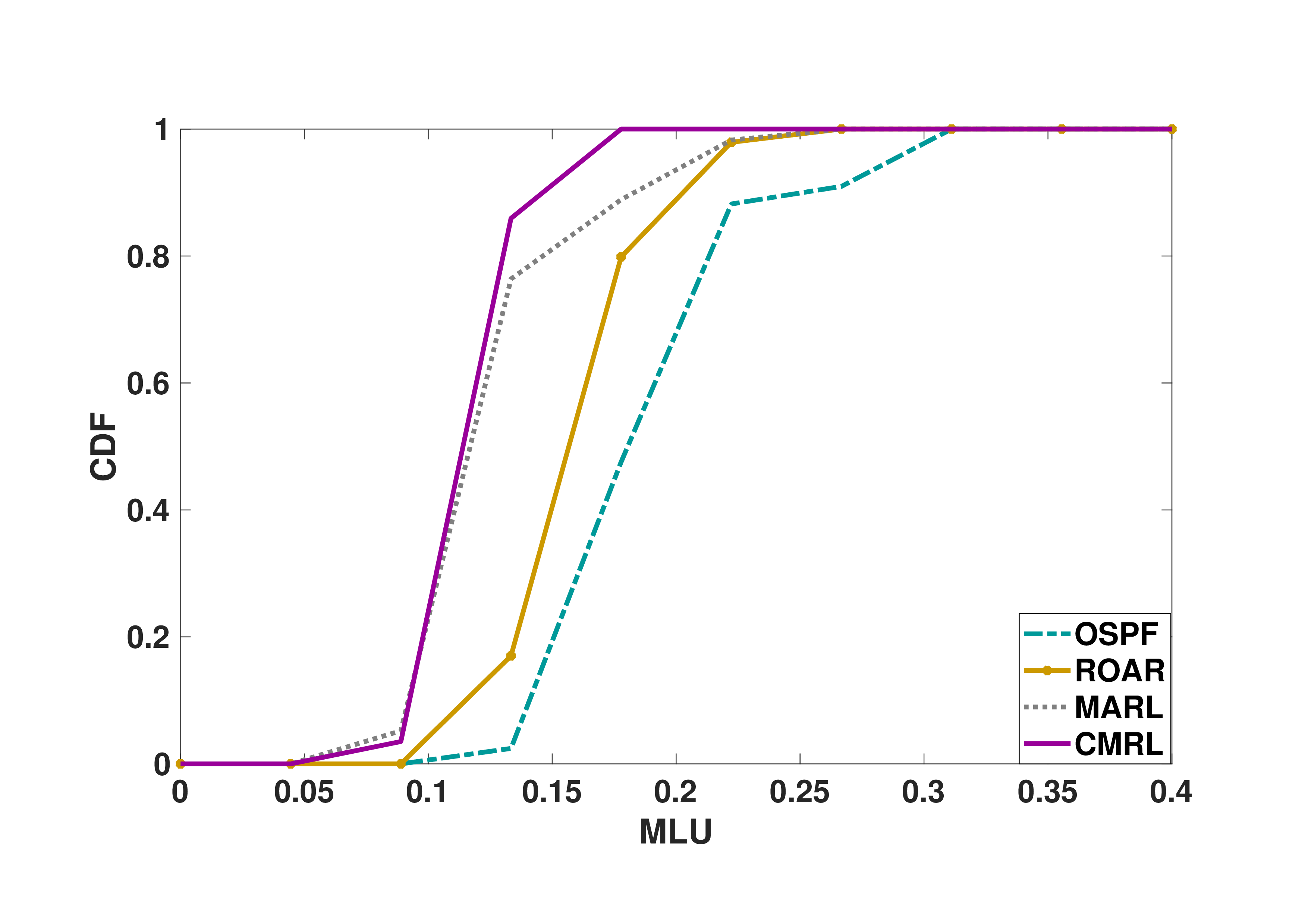}
}
\subfloat[0.4]{
\includegraphics[width=0.3\textwidth]{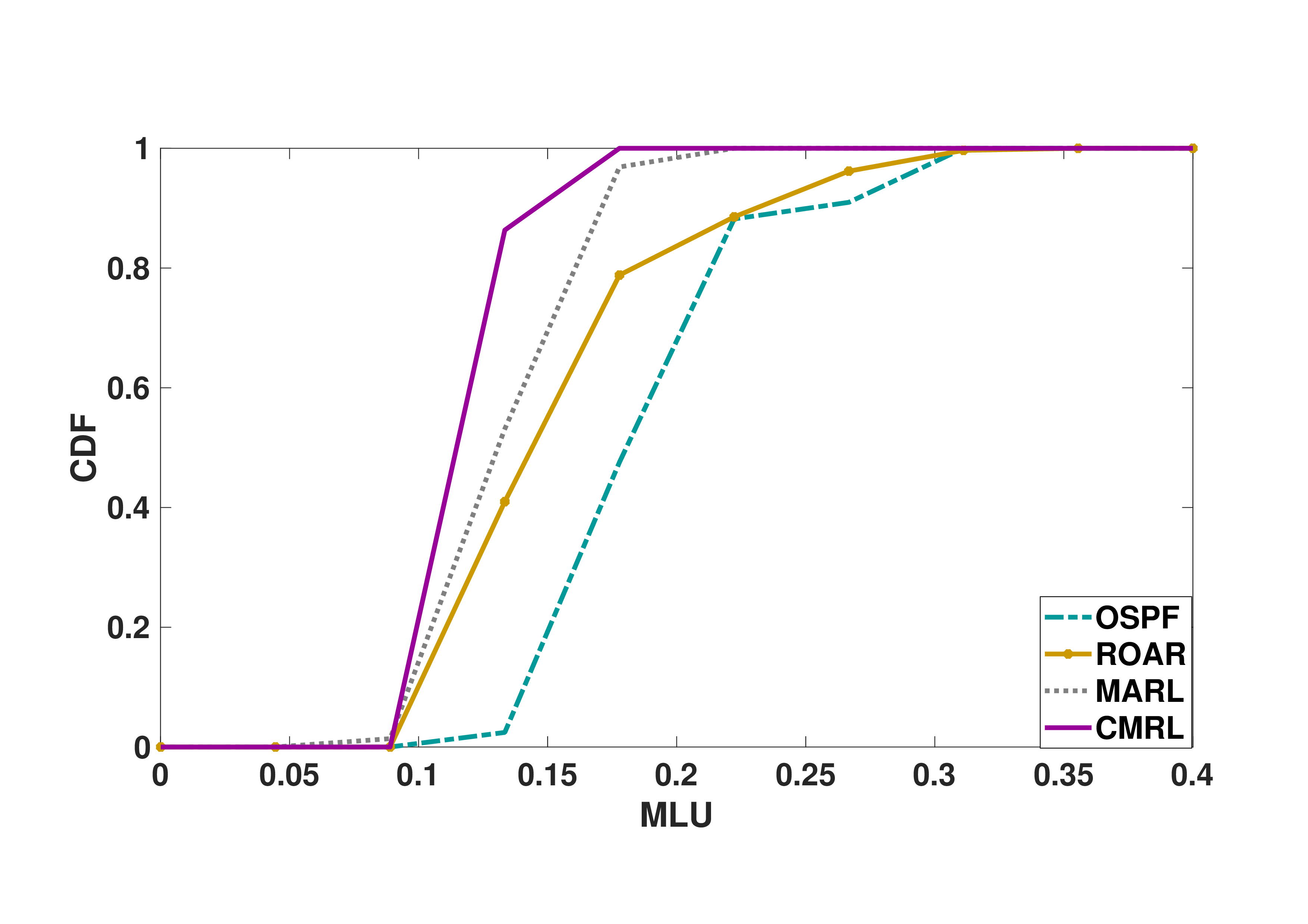}
}
\vspace{-0.2cm} 
\caption{The CDF curves of MLU under different SDN deployment ratios in G\'{E}ANT. }
\label{geant}
\end{figure*}

%
%
%
%

\subsubsection{Network performance under dynamic traffic demands}

\begin{figure*}[!htb]
\vspace{-0.1cm} 
\centering
\subfloat[Abilene]{
\includegraphics[width=0.3\textwidth]{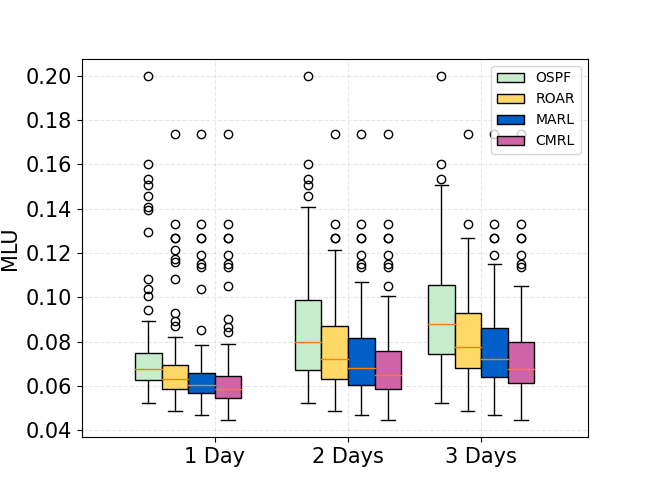}
}
\subfloat[CERNET]{
\includegraphics[width=0.3\textwidth]{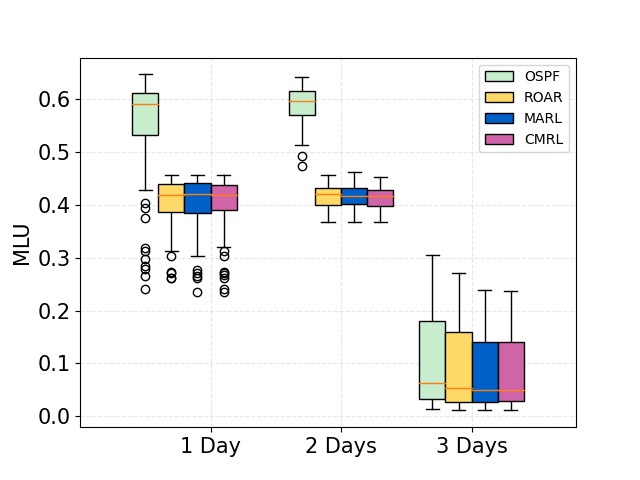}
}
\subfloat[G\'{E}ANT]{
\includegraphics[width=0.3\textwidth]{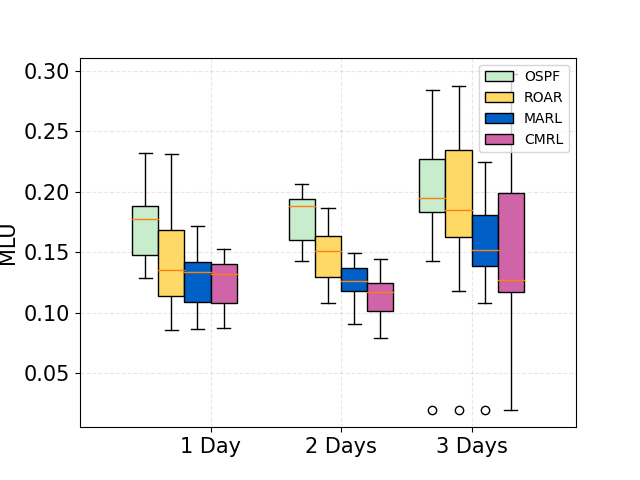}
}
\caption{The MLU under different time periods. We evaluate the performance under 288TMs (1 day), 576TMs (2 days) and 864TMs (3 days) for Abilene and CERNET topologies. We evaluate the performance under 96TMs (1 day), 192TMs (2 days) and 288TMs (3 days) for G\'{E}ANT topology.  }
\vspace{0.2cm} 
\label{day}
\end{figure*}

\begin{figure*}[!htb]
\vspace{-0.5cm} 
\centering
\subfloat[Abilene]{
\includegraphics[width=0.3\textwidth]{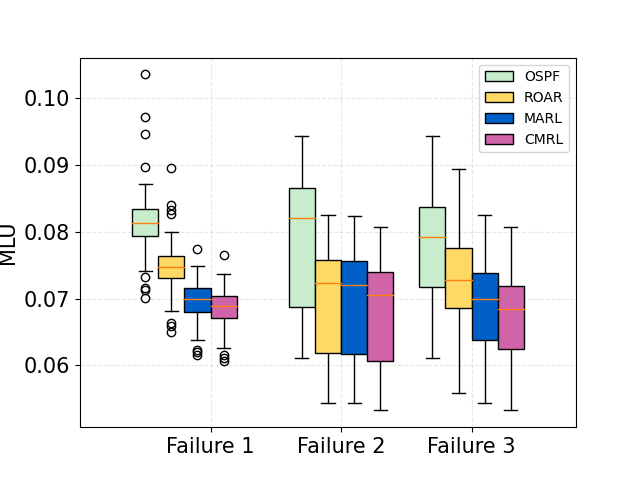}
}
\subfloat[CERNET]{
\includegraphics[width=0.3\textwidth]{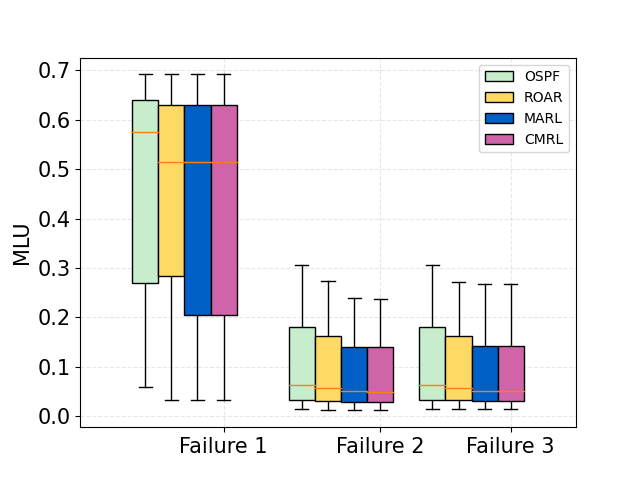}
}
\subfloat[G\'{E}ANT]{
\includegraphics[width=0.3\textwidth]{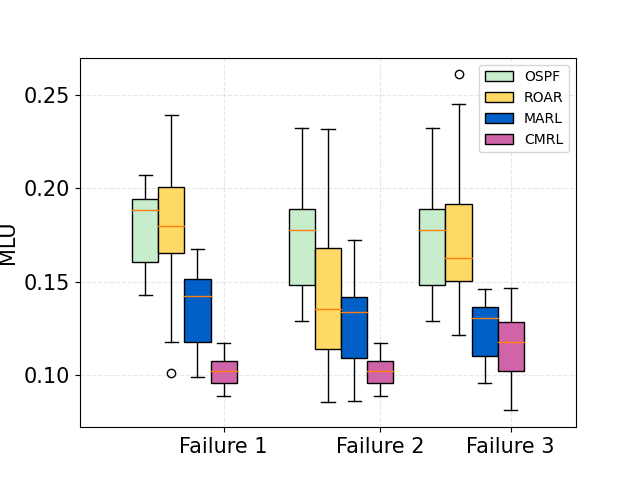}
}
\caption{The MLU of network failure scenarios in different network topologies. }
\vspace{-0.6cm}
\label{failure}
\end{figure*}

To comprehensively evaluate the TE performance of various methods under dynamic traffic demands, we plot the Cumulative Distribution Function (CDF) curves of MLU under different SDN deployment ratios and different network topologies in Fig. \ref{ab}, Fig. \ref{cernet} and Fig. \ref{geant}. As shown in Figs. \ref{ab}-\ref{geant}, we can observe that the CDF curves of CMRL stay above the CDF curves of OSPF, ROAR and MARL under Abilene, CERNET and GEANT network topologies with SDN deployment ratios set to $0.2$, $0.3$ and $0.4$, respectively. This demonstrates that compared to the MLU derived in other methods, our proposed method CMRL can obtain a lower MLU under different network topologies and SDN deployment ratios. In addition, with the increasing of network scale, the gap between CMFL and the other methods becomes larger. This is because in large network topologies, there are more adjacent links for each switch and the agents have more flexibility in optimizing the distribution of flows through the SDN switch.\par

The reason why CMRL outperforms OSPF in minimizing MLU is that the traffic is constrained to routed on the shortest paths between the source-destination node pair in OSPF, while in our method CMFL, traffic can be split on multiple available paths from the source to the destination, thus achieving better link load balance. Meanwhile, compared to single-agent RL method ROAR, our method has superior performance. The reason is the collaboration of multiple agents can better learn the mapping between traffic demands and routing policies. When new traffic demands arrive, the trained agents can better infer the routing policies with a higher TE performance. Additionally, compared to general MARL, the TE performance of CMRL can be greatly improved with the introduction of difference reward assignments, which encourages multiple agents to take better actions. \par

\begin{table}[!htb]
\renewcommand\arraystretch{1.2}
\centering
\caption{Average MLU under different traffic demands }\label{mlu}
\begin{tabular}{|p{1cm}|p{2cm}|p{2.1cm}|p{2.2cm}| } \hline
Method & Abilene&CERNET&G\'{E}ANT  \\ \hline
OSPF    &0.124(32.50\%)&0.108(19.35\%) &0.182 (41.21\%)     \\ \hline
ROAR    &0.0848(1.29\%)&0.0973(10.48\%)&0.146 (26.71\%)         \\ \hline
MARL    &0.0847(1.18\%)&0.0989(11.93\%)&0.140 (23.57\%) \\ \hline
CMRL    &0.0837&0.0871&0.107 \\ \hline
\end{tabular}
\vspace{0.2cm}
\end{table}

To exhibit the concrete performance improvement ratio of CMRL, we compute the average MLU of different algorithms under different network topologies in TABLE \ref{mlu}. As shown in TABLE \ref{mlu}, we can observe that compared to OSPF, our proposed method CMFL improves the network performance up to $41.21\%$ under different network topologies; compared to ROAR, our proposed method improves the network performance up to $26.71\%$ under different network topologies; compared to MARL, our proposed method improves the network performance up to $23.57\%$ under different network topologies.\par

In addition, we conduct extensive experiments to comprehensively evaluate the TE performance under different number of TMs in Fig. \ref{day}. Each box contains the $5\%$-quantile, $25\%$-quantile, median value, $75\%$-quantile and the outliers. As shown in Fig. \ref{day}(a), we can observe that compared to other methods, our proposed method CMRL obtains a smaller $5\%$-quantile, $25\%$-quantile, median value, $75\%$ MLU value under TMs of different time periods in Abilene topology. The results are similar in CERNET and G\'{E}ANT topologies. Through the extensive experiments, we can conclude that CMRL can efficiently learn the routing policies through the collaboration of multiple agents and the network performance can be better enhanced in CMRL, compared with other TE methods.\par

\subsubsection{TE performance under network failures}

\begin{table}[!htb]
\renewcommand\arraystretch{1.2}
\vspace{0.2cm}
\centering
\caption{Average MLU under link failure scenarios }\label{table:time}
\begin{tabular}{|p{1cm}|p{2.1cm}|p{2.1cm}|p{2.0cm}|} \hline
Method & Abilene&CERNET&G\'{E}ANT  \\ \hline
OSPF    &0.0790(14.94\%)&0.460(7.83\%)&0.173(32.95\%)       \\ \hline
ROAR    &0.0689(24.67\%)&0.442(4.07\%)&0.170(31.76\%)      \\ \hline
MARL    &0.0687(2.18\%)&0.430(1.40\%)&0.125(7.20\%)         \\ \hline
CMRL    &0.0672&0.424&0.116         \\ \hline
\end{tabular}
\vspace{-0.2cm}
\end{table}

Network failures, especially single link failures, happen frequently in large networks and can lead to severe network congestion and packet loss \cite{liu2014traffic}. To validate the superior performance of CMRL in handling network failures, we evaluate the MLU of the hybrid SDN under different single link failures and draw the box diagrams in Fig. \ref{failure}. As shown in Fig. \ref{failure}(a), we can observe that under different link failures, our proposed method CMRL can obtain a smaller MLU compared to the MLU derived from the other methods in Abilene. In CERNET and G\'{E}ANT shown in Fig. \ref{failure}(b) and \ref{failure}(c), we obtain the similar results. In particular, the gap between our method CMFL and other baseline methods becomes larger in the G\'{E}ANT network. In summary, our proposed method CMFL exhibits superior performance in minimizing MLU of the hybrid SDN. As shown in TABLE \ref{table:time}, our proposed method can reduce the average MLU up to  $32.95\%$, $31.76\%$ and $7.20\%$ compared to OSPF, ROAR and MARL, respectively, under different network topologies. This demonstrates that CMRL is robust to network failures and can infer an efficient routing policy when link failures happen.


\subsubsection{Online inference time}

Finally, we record the online inference time of different methods in TABLE \ref{time}. As shown in TABLE \ref{time}, we can observe that CMRL has a shorter online inference time compared to ROAR and the inference time of CMRL approximates that of MARL. This demonstrates that compared to ROAR, the computation time can be reduced in CMRL by dividing a large-scale inference problem into multiple small-scale inference problems. The trained agents in CMRL can promptly react to the dynamic changing environment in an online manner when traffic demands change or network failures happen.

\begin{table}[!htb]
\renewcommand\arraystretch{1.2}
\centering
\caption{The online inference time }\label{time}
\begin{tabular}{|c|c|c|c|} \hline
Method &Abilene&CERNET &G\'{E}ANT   \\ \hline
ROAR   &0.878ms &0.891ms &1.203ms           \\ \hline
MARL   &0.534ms  &0.547ms & 0.996ms   \\ \hline
CMRL     &0.531ms &0.545ms   &0.996ms\\ \hline
\end{tabular}
\vspace{-0.4cm}
\end{table}


\section{Conclusion}\label{s6}
In this paper, we innovatively propose a multi-agent reinforcement learning framework CMRL for improving the TE performance in a dynamic hybrid SDN environment. Specifically, an interactive environment is first constructed and the multiple agents are trained offline for collaboratively learning the map between traffic demands and routing policies. To solve the credit assignment of multi-agent, difference reward assignment is introduced for encouraging the agents to sacrifice for the good actions. Then, the trained agents are deployed for online routing policy inference. The extensive experiments demonstrate the superior performance of CMRL in reducing MLU of hybrid SDN when traffic demands change or network failures happen.

\section{Acknowledgements}\label{Acknowledgements}
This work is partially supported by National Natural Science Foundation of China under Grant No.62002064, and the Natural Science Foundation of Fujian Province under Grant 2020J05110.

\bibliographystyle{IEEEtran}
\bibliography{DMAC}

\end{document}